\documentclass[aps,draft]{revtex4}
\usepackage[dvips]{graphicx}
\begin{document}

\title{ A new Concept of Ball Lightning }
\author{Levan N.Tsintsadze}
\thanks{Also at Department of Plasma Physics, E.Andronikashvii Institute of Physics, Tbilisi,
Georgia}
\affiliation{Graduate School of Pure and Applied Sciences, University of Tsukuba, Tsukuba, Japan}

\date{\today}

\begin{abstract}
We suggest that the ball lightning (BL) is a weakly ionized gas, in which the electromagnetic radiation can be
accumulated through the Bose-Einstein condensation and/or the photon trapping in the plasma density well. We derive the set of
equations describing the stability of BL, and show that the BL moving along charged surface becomes unstable. Eventually the instability leads to explosion of BL and release of energy of the trapped photons and/or the Bose-Einstein condensate.
\end{abstract}


\maketitle

Ball lightning (BL) is one of the extraordinary and least understood atmospheric phenomena, which appears infrequently and unpredictably. The phenomenon lasts only a short time, before either fading away or violently dissipating with an explosion. Scientists have postulated that plasma may be behind the phenomenon. They have also discussed links between some lightning events and mysterious gamma ray emissions that emanate from earth's own atmosphere,  suggesting that this gamma radiation fountains upward from starting points surprisingly low in thunderclouds. An early attempt to explain BL was recorded by Tesla in 1904, however currently, there is no widely accepted explanation of what exactly BL is.

In this letter, we present our interpretation of BL. There is an evidence from the observations that during the thunderstorm the linear discharge of an atmosphere bends significantly in some place, occasionally bends as a loop. In such cases the self-magnetic field pressure inside of the curve is rather large than outside of it. This implies that the magnetic force can eject a plasma bunch with certain velocity in the atmosphere. We here assume that the plasma bunch, in which plasma is considered to be incompressible, has a spherical form with radius $R_0$. We suggest that the plasma ball or ball lightning is a weakly ionized gas. That is collisions between electrons (ions) and molecules are more important than the electron-electron and electron-ion collisions, so that the latter may be neglected. We shall also assume that the mean energy acquired by the electrons in the electric field during the atmospheric discharges is insufficient to ionize or excite the molecules. Hence, the collisions between electrons (ions) and molecules may be regarded as elastic.
The frictional force we then can write as \cite{gin}
\begin{eqnarray}
\label{ff}
R_\alpha=\sum_\beta\int\vec{v}_\alpha C_{\alpha \beta }d\vec{v}=m_\alpha
n_\alpha\sum_\beta\nu_{\alpha\beta }(v_\alpha - v_\beta)
\end{eqnarray}
where $\nu_{\alpha \beta }$ is the collision frequency of particles $\alpha$ with particles $\beta $, $m_\alpha$ and $n_\alpha$ are mass and densities of particles, respectively.

In the study of BL one of the most salient and complex question is how the radiation accumulates in the BL. However, to the best of our knowledge the previous investigations did not elucidate a mechanism of accumulation. Thus, the mechanism of accumulation of radiation in the BL remains unsolved issue of the high importance. Here we suggest a new concept of the accumulation of radiation through two possible mechanisms. One is the accumulation of a strong radiation via Bose-Einstein (BE) condensation, which was recently predicted by Tsintsadze \cite{ltsin02} demonstrating that the life time of BE condensate is rather large before start of diffusion of photons from the condensate state.

The second mechanism of the accumulation of radiation is due to the density inhomogeneity of BL plasma. So that the existence of strong deviation of the density ($\delta n_{p}<0$) of plasma from the equilibrium due to inhomogeneity, can lead to the creation of a potential well for photons. Simple calculations show that all photons with the wave number $k<k_p=\frac{\omega_p}{c},$ will be trapped in the potential well. In this case for the number density of the trapped photons one obtains \cite{ltsin04}
\begin{eqnarray}
\label{ndtp}
n_\gamma^{trap}=\frac{k_p^3}{3\pi^2}\mid\frac{\delta n_p}{n_{0p}}\mid^{3/2}\ ,
\end{eqnarray}
where $\omega_p=\left( \frac{4\pi e^2n_{0p}}{m_e}\right)^{1/2}$ is the Langmuir frequency.

For the energy of a photon, which is trapped in the well, we can write
\begin{eqnarray}
\label{pen}
U_{trap}=\frac{\hbar\omega_p}{10\pi^2}k_p^3V\mid\frac{
\delta n_p}{n_{0p}}\mid^{5/2}
\end{eqnarray}
where $V=\frac{4\pi}{3}R_0^3$ is the volume of BL.

We now can define the total energy of the photon gas, which is accumulated in
the BL, as
\begin{eqnarray}
\label{ten}
E=N_{trap}U_{trap}=\frac{\hbar\omega_p}{30\pi^4}(k_p^3V)^2\mid\frac{
\delta n_p}{n_{0p}}\mid^4 \ .
\end{eqnarray}
Note that all other photons, with the wave number larger than $k_p$, are untrapped
and they can freely leave the BL.

In order to describe a plasma that consists of electrons, ions and molecules
in BL, we employ the following set of equations \cite{ltsin95}
\begin{eqnarray}
\label{setc}
\frac{\partial n_\alpha}{\partial t}+div(n_\alpha
\vec{v}_\alpha )=0 \ ,
\end{eqnarray}
\begin{eqnarray}
\label{setm}
n_\alpha\frac{dp_{\alpha j}}{dt}=e_{\alpha }n_{\alpha }(\vec{E}
+\frac{1}{c}\vec{v}_\alpha\times\vec{B})_j-\nabla_jP_\alpha+R_{\alpha j}+m_\alpha n_\alpha g_j\ ,
\end{eqnarray}
\begin{eqnarray}
\label{setp}
div\vec{E}=4\pi\rho_e \ ,
\end{eqnarray}
where $n_\alpha ,v_\alpha ,P_\alpha ,p_{\alpha j}$ are the number density, the mean velocity, the
pressure, the mean momentum, respectively, of each species, $\rho_e=e(n_i-n_e)$, $g$ is
the gravitational acceleration, and the ion charge is assumed to be $e$, i.e., $Z_i=1$.

Taking into account the conservation of momentum, i.e., $\sum_\alpha R_{\alpha j}=0$, we get from Eq.(\ref{setm})
\begin{eqnarray}
\label{mom}
\sum_\alpha n_\alpha\frac{dp_{\alpha j}}{dt}=\rho_e\vec{E}+
\frac{1}{c}\vec{j}\times\vec{B}-\nabla_j\sum_\alpha P_\alpha+\sum_\alpha m_\alpha n_\alpha\vec{g} \ ,
\end{eqnarray}
where $\vec{j}=-e(n_e\vec{v}_e-n_i\vec{v}_i)$.

Since we consider the weakly ionized dense plasma the collision frequency of electrons $\nu _{eN}$ (ions $\nu _{iN}$) with molecules is essentially larger than $\tau^{-1}$ ($\nu_{\alpha N}\tau >>1$), where $\tau$ is the characteristic time during which the plasma parameters change significantly. Furthermore, the mean velocities of different components of the plasma must be almost equal $\left( \vec{v}_e\simeq\vec{v}_i\simeq\vec{v}_N\right)$, i.e., only in this case can the frictional forces be balanced by the other terms in Eq.(\ref{setm}). With these conditions at hand Eq.(\ref{mom}) is reduced to one-fluid equation
\begin{eqnarray}
\label{onef}
\rho_M\frac{d\vec{v}}{dt}=\rho_e\vec{E}-\nabla\left(P+\frac{B^2}{8\pi}\right)+\frac{1}{4\pi}
(\vec{B}\cdot\nabla)\vec{B}+\rho_M\vec{g} \ ,
\end{eqnarray}
where use was made of the Maxwell equation $\vec{j}=\frac{c}{4\pi}curl\vec{B}$,
$\rho_M=m_en_e+m_in_i+m_Nn_N$ is the mass density of three component plasma, and
$P=\sum_\alpha P_\alpha$.

Equation (\ref{setc}) is also rewritten as
\begin{eqnarray}
\label{con}
\frac{\partial \rho_M}{\partial t}+div(\rho_M
\vec{v})=0 \ .
\end{eqnarray}

We now apply these Eqs.(\ref{onef}),(\ref{con}) to the surface of BL. Note that
the tangential components of the field $\vec{E}$ at the surface
of BL, as conductor, must be zero, i.e., $\vec{E}_t=0$ and the
electrostatic field $\vec{E}=-\nabla\varphi$ must be normal to the surface at
every point \cite{lanel}. Thus, we can write
\begin{eqnarray}
\label{nef}
\left(\rho_e\vec{E}\right)_{\vec{n}}=\frac{1}{4\pi}(\vec{E}div\vec{E})_n=\frac{1}{8\pi}\frac{
\partial}{\partial\vec{n}}E^2 \ ,
\end{eqnarray}
where the vector $\vec{n}$ is the normal of the surface and $\vec{E}=-\frac{\partial\varphi}{\partial\vec{n}}$.

It should be emphasized that the term on the right-hand side of Eq.(\ref{nef}), $\frac{E^2}{8\pi}$ is the negative pressure, which acts on the charged surface of an $e^{-}+i+N$ plasma. We also recall that the plasma is supposed to be
incompressible. Then, the equation of continuity (\ref{con}) takes the simple form for constant $n_e+n_i+n_N$
\begin{eqnarray}
\label{simc}
div\vec{v}=0 \hspace{1cm} or \hspace{1cm} \nabla^2\Psi=0 \ ,
\end{eqnarray}
where $\Psi$ is the potential of velocity, i.e., $\vec{v}=\nabla\Psi$.

We specifically note here that a steady-state force balance equation on the surface can be deduces from
Eq.(\ref{onef}) at $\vec{v}=0$ and $(\vec{B}\cdot\nabla)\vec{B}=0$.

We now consider the stability of BL, assuming that at first BL has a spherical form. The velocity potential $\Psi$ satisfies
Laplace's equation (\ref{simc}) with boundary condition $r=R_0$, the solution of which is well known and can be represented as $\Psi =g(r)r^lY_{l,m}(\theta ,\varphi )$. Here $Y_{l,m}(\theta ,\varphi )$ are Laplace's spherical harmonics, which satisfy the following equation \cite{lanq}
\begin{eqnarray}
\label{lfeq}
\frac{1}{\sin \theta }\frac{\partial }{\partial \theta }\left(\sin \theta \frac{
\partial Y_{l,m}(\theta ,\varphi )}{\partial \theta }\right)+\frac{1}{\sin^2\theta }\frac{\partial^2}{
\partial\varphi^2}Y_{l,m}+l(l+1)Y_{l,m}=0 \ ,
\end{eqnarray}
where $l=0,1,2,3...$
Thus along the radius of the sphere the component of velocity $v_r=\frac{\partial\Psi}{\partial r}=\frac{l\Psi}{r}$
will be zero when $l=0$.

As mentioned above, we suppose that in equilibrium the surface of BL is spherical. Under the action of some perturbation, there will be propagated waves over the whole surface. In order to investigate the stability of BL, we shall derive a dispersion relation of the surface waves. To this end, we can linearize Eqs.(\ref{onef}) and (\ref{nef}), because the amplitude of perturbation is small. That is we assume that the shape of sphere little changes. We note also that the magnetic field $\vec{B}$ is the self-magnetic field, which is generated during the linear discharge, and after linearization disappears in equation.

In equilibrium, when the surface is at rest, the electric field above it, as
follows from Eq.(\ref{setp}) (using the fact that $\vec{E}=0$ on the
inner area), takes the form
\begin{eqnarray}
\label{pun}
E_{\vec{n}}=4\pi\int\rho_ed\vec{n}=4\pi\sigma \ ,
\end{eqnarray}
where $\sigma$ is the surface charge density.

From expression (\ref{pun}), it is
clear that we have for the potential $\Phi$,
\begin{eqnarray}
\label{pot}
\Phi =-4\pi\sigma r \ .
\end{eqnarray}
On the surface itself the potential must have a constant value. Then the area of surface given in spherical polar coordinates $r,\theta ,\varphi $ can be expressed by a function $r=r(\theta ,\varphi,t)$. When the surface suffers some perturbation, the potential (\ref{pot}) above the oscillating surface can be written as
\begin{eqnarray}
\label{apot}
\Phi =-4\pi\sigma (R_0+\zeta (\theta ,\varphi,t))+\delta\Phi
\end{eqnarray}
since the spherical surface is given by $r=R_0$ and neighboring surface by
$r=R_0+\zeta$, and we have assumed that $\zeta (\theta ,\varphi ,t)$ is everywhere
small, i.e., the surface deviates only slightly from $R_0$.

Smallness of $\zeta$ allows us to express the radial velocity $v_r$ through
the $\zeta(\theta,\varphi,t)$ as $v_r=\frac{\partial \zeta }{\partial t}$. On the other hand $v_r=\frac{\partial\Psi }{
\partial r}$, so that we can write
\begin{eqnarray}
\label{sim}
\frac{\partial\zeta}{\partial t}=\frac{\partial\Psi}{\partial r}\ .
\end{eqnarray}
Also from Eq.(\ref{apot}) follows that $\delta\Phi=4\pi\sigma\zeta$.

Furthermore, since the oscillations are small, we have
\begin{eqnarray}
\label{fiel}
\frac{E_n^2}{8\pi}=\frac{E_0^2}{8\pi}+\frac{E_0\delta E}{4\pi}=2\pi\sigma^2-\sigma\frac{\partial\delta\Phi}{\partial r}\ .
\end{eqnarray}
The potential distribution satisfies the equation $\Delta\delta\Phi=0$
with the boundary condition $r=R_0$. Therefor, solving the Laplace's equation for $\delta\Phi$
\begin{eqnarray}
\label{lapl}
\frac{1}{r^2}\frac{\partial}{\partial r}r^2\frac{\partial}{\partial r}\delta\Phi -\frac{l(l+1)}{r^2}\delta\Phi=0
\end{eqnarray}
we find the solution that vanishes in the limit $r\rightarrow\infty$, which is $\delta\Phi =\frac{a(t)}{r^{l+1}}$ or
\begin{eqnarray}
\label{lsol}
\delta E=\frac{(l+1)}{r}\delta\Phi =\frac{4\pi\sigma (l+1)}{r}\zeta \ .
\end{eqnarray}
The equation of motion (\ref{onef}), after linearization, for the radial
component of the velocity of points on the surface reads
\begin{eqnarray}
\label{lmot}
\left(\frac{\partial v_r}{\partial t}+\frac{1}{\rho_M}\frac{\partial}{
\partial r}\delta P-\frac{1}{\rho_M}4\pi\sigma^2(l+1)\frac{\partial}{
\partial r}\frac{\delta\Phi}{r}+g\right)_{r=R_0}=0\ ,
\end{eqnarray}
where $\delta P$ is the pressure difference between the two sides of the
surface and is not zero, which is given by Laplace's formula \cite{lanf} in the
spherical polar coordinates
\begin{eqnarray}
\label{pres}
\delta P=P_1-P_2=\alpha\left(\frac{1}{R_1}+\frac{1}{R_2}\right)=
\alpha\left\{\frac{2}{R_0}-\frac{2\zeta}{R_0^2}-\frac{1}{R_0^2}
\left[\frac{1}{\sin^2\theta}\frac{\partial^2\zeta}{\partial\varphi^2}+
\frac{1}{\sin\theta}\frac{\partial}{\partial\theta}\sin\theta
\frac{\partial}{\partial\theta}\zeta\right]\right\}= \nonumber \\
\alpha\left\{\frac{2}{R_0}-\frac{2\zeta}{R_0^2}+\frac{l(l+1)}{R_0^2}\zeta\right\}\ ,
\end{eqnarray}
where $\alpha $ is the surface-tension coefficient.

Substituting $v_r=\frac{\partial\Psi}{\partial r}$ and Eq.(\ref{pres}) into Eq.(\ref{lmot}), and integrating
Eq.(\ref{lmot}) over  $r$, and differentiating with respect to time, using
Eqs.(\ref{sim}),(\ref{lsol}) and $\frac{\partial\Psi}{\partial r}=\frac{l\Psi}{r}$, we arrive at the equation for the potential of velocity
\begin{eqnarray}
\label{potve}
\frac{\partial^2\Psi}{\partial t^2}+\frac{1}{R_0}\left[\frac{\alpha}{\rho_MR_0^2}(l-1)(l+2)+g-\frac{4\pi\sigma^2(l+1)}{
\rho_MR_0}\right]\Psi_{\mid_{r=R_0}}=0 \ .
\end{eqnarray}
We shall look for a solution which is a simple function of time $\Psi =f\exp(-i\omega t)$, and substitute it into
Eq.(\ref{potve}) to obtain the dispersion relation for the surface waves (electrostatic capillary gravity waves), the result is
\begin{eqnarray}
\label{disp}
\omega^2=\frac{1}{R_0}\left\{\frac{\alpha}{\rho_MR_0^2}
(l-1)(l+2)+g-\frac{4\pi\sigma^2(l+1)}{\rho_MR_0}\right\}\ .
\end{eqnarray}
Important conclusions are evident from this dispersion relation. Namely, if the surface of BL is not charged, then BL remains stable ($\omega^2>0$) even when the shape of BL changes. However, in the presence of charge on the surface, for the certain density $\sigma$ of surface charge, $\omega^2$ can be negative, i.e., BL becomes unstable, the amplitude of waves exponentially grows and in the nonlinear stage BL will explode.

If we analyze Eq.(\ref{disp}), we can see that the capillary oscillations are absent for $l=1$, as in this case
\begin{eqnarray}
\label{capil}
\omega^2=\frac{1}{R_0}\left( g-\frac{8\pi\sigma^2}{\rho_MR_0}\right)\ .
\end{eqnarray}
One can estimate $\sigma$ from this simple relation. The dispersion relation (\ref{capil}) implies that the instability arises for $\sigma >\sqrt{\frac{\rho_MR_0}{8\pi}g}$.
Whereas a general condition of the instability is
\begin{eqnarray*}
4\pi\sigma^2=\frac{E_0^2}{4\pi}>\frac{\rho gR_0}{l+1}+\frac{\alpha}{R_0}\frac{(l-1)(l+2)}{l+1}\ .
\end{eqnarray*}
From here it is clear that the role of the capillary waves increases with increase of $l$, and for some $l_0$ the plasma becomes stable. Note that this parameter $l_0$ depends on the amount of charge on the surface.

To summarize, we have proposed a new concept for BL. Namely, that BL is a weakly ionized gas, in which the electromagnetic radiation can be accumulated through the Bose-Einstein condensation \cite{ltsin02} and/or the photon trapping in the plasma density well \cite{ltsin04}. We have derived the set of equations describing the stability of BL, and have shown that the BL moving along any uncharged surface remains stable, even when the shape of BL changes, and it can penetrate into any split, hole, chimney, etc. However, if the surface is charged, on which the BL moves, then the surface of BL is also charged. In this case, the surface of BL becomes unstable \cite{ltsin98}. Eventually the instability leads to explosion of BL and release of energy of the trapped photons and/or the Bose-Einstein condensate.

\end{document}